\documentclass[aps,prl,twocolumn,superscriptaddress,showpacs]{revtex4}

\usepackage{graphicx}
\usepackage{dcolumn}
\usepackage{bm}

\begin{document}

\preprint{APS/123-QED}

\title{First Experimental Observation of Superscars in a Pseudointegrable Barrier Billiard}

\author{E.~Bogomolny}
\affiliation{Laboratoire de Physique Th{\'e}orique et Mod{\`e}les Statistiques, Universit{\'e} de Paris-Sud, B{\^a}timent 100, 91405 Orsay Cedex, France}

\author{B.~Dietz}
\affiliation{Institut f{\"u}r Kernphysik, Technische Universit{\"a}t
Darmstadt, D-64289 Darmstadt, Germany}

\author{T.~Friedrich}
\affiliation{Institut f{\"u}r Kernphysik, Technische Universit{\"a}t
Darmstadt, D-64289 Darmstadt, Germany}

\author{M.~Miski-Oglu}
\affiliation{Institut f{\"u}r Kernphysik, Technische Universit{\"a}t
Darmstadt, D-64289 Darmstadt, Germany}

\author{A.~Richter}
\email{richter@ikp.tu-darmstadt.de}
\affiliation{Institut f{\"u}r
Kernphysik, Technische Universit{\"a}t Darmstadt, D-64289 Darmstadt,
Germany}

\author{F.~Sch{\"a}fer}
\affiliation{Institut f{\"u}r Kernphysik, Technische Universit{\"a}t
Darmstadt, D-64289 Darmstadt, Germany}

\author{C.~Schmit}
\affiliation{Laboratoire de Physique Th{\'e}orique et Mod{\`e}les Statistiques, Universit{\'e} de Paris-Sud, B{\^a}timent 100, 91405 Orsay Cedex, France}
\date{\today}

\begin{abstract}
With a perturbation body technique intensity distributions of the electric field strength 
in a flat microwave billiard with a barrier inside up to mode numbers as large as about 700 were measured. A method for the reconstruction of the amplitudes and phases of the electric field strength from those intensity distributions has been developed. 
Recently predicted superscars have been identified experimentally and --- using the well known analogy between the electric field strength and the quantum mechanical wave function in a two-dimensional microwave billiard --- their properties determined.

\end{abstract}

\maketitle


 Planar polygonal billiards with angles $\alpha_j=\pi m_j/n_j$, where $m_j$ and $n_j$ are co-prime integers, have been studied both
classically and quantum mechanically \cite{RichensBerry1980}. When $m_j\ne 1$ the motion in phase space is not restricted to a torus like for integrable systems, but 
to a surface with a more complicated topology. Accordingly, such planar polygonal billiards are 
called pseudointegrable. It was established numerically (see \cite{Bogomol1999,Wiersig2002} and refs.\
therein) that the statistical properties of the eigenvalues of the corresponding quantum systems are intermediate between those of a regular and a chaotic system. The properties of the wave functions of planar polygonal billiards are also intriguing, as they show a
strong scarring behavior  which can be related to families of 
periodic orbits \cite{Bogomol2004}. In a plot of eigenfunctions 
in the barrier 
billiard scars are clearly distinguishable from non-scarred eigenfunctions.
This pronounced scar structure
does not disappear at large quantum numbers in contrast to that in chaotic 
systems \cite{Kaufmann83,Heller1984,Antonsen95}. To stress this
difference it is proposed in \cite{Bogomol2004} to call the scars in pseudointegrable systems superscars, an expression used by Heller in his early seminal paper \cite{Heller1984} in a different context.
The aim of this letter is to report on the experimental investigation of superscars in the barrier billiard. This is a rectangular billiard of area $l_x \cdot l_y$ which contains an
infinitely thin barrier. In the experiment presented here the barrier is placed on the symmetry line $x=l_x/2$
and its length equals $l_y/2$, where $l_y$ is the length of the shorter side of the rectangle. 

\par In the present work the quantum barrier billiard is simulated by means of a microwave billiard.
Microwave billiards are flat cylindrical resonators \cite{Stoeckmann:buch00,richter97}.
Below the critical frequency $f_c=c/2h$, where $c$ is the velocity of light and $h$ is the height of the
cavity, the electric field is the solution of the scalar Helmholtz
equation with Dirichlet boundary conditions. This equation is mathematically equivalent to the Schr{\"o}dinger equation for a quantum billiard of corresponding shape (see e.g.\ \cite{Stoeckmann:buch00,richter97}).

\par The experimental observation of the superscars predicted in \cite{Bogomol2004} 
requires (i) the detection of eigenmodes at sufficiently high quantum numbers $N$ and (ii) the resolution of all modes up to these quantum numbers. The number $N(f_c)$ of resonances below the critical frequency $f_c$ is
approximately given by the first term of the Weyl formula \cite{Weyl1912}, i.e.\ $N(f_c)=\frac{A}{4\pi} \left( \frac{2\pi}{c}\,f_c\right) ^2 =\frac{A}{4\pi} \left( \frac{\pi}{h}\right) ^2,$ where $A$ is the area of the billiard. 
Hence, the number of experimentally accessible resonances increases with the area and decreases with
the height of the billiard. The resonances can be well resolved, if their widths are small in comparison with the average
spacing between adjacent resonances, i.e.\ a microwave cavity with a high quality factor $Q$ is needed. The latter is proportional to the ratio of the volume to the 
surface of the cavity, therefore for flat cylindrical resonators to the height of the
resonator. A compromise between a high $Q$ and a large number of resonances is obtained by designing a rectangular cavity with side lengths of $l_x=1000~\rm{mm}$ and $l_y=585~\rm{mm}$, and a height of $h=25~\rm{mm}$. The barrier has  length  297~mm, a width of 4~mm and the same height as the rectangular cavity. 
As in our earlier experiments \cite{Dembowski2000} the resonator consists of a bottom, a lid and a frame in between. All parts are squeezed together with screws. The bottom and lid plates respectively are manufactured from polished copper. The frame has the shape of a rectangle and consists of 10 brass segments. To ensure a proper electrical contact even at high frequencies, wires of solder are placed between the bottom, the frame and the lid. The so constructed microwave barrier billiard has about 700 resonances below $f_c=6~\rm{GHz}$, the average quality factor is $Q=1.8\cdot10^4$ and the mean resonance width is at least 10 times smaller than the mean resonance spacing.
\par In order to obtain information on the wave functions in the quantum barrier billiard we determined the electric field strength distribution in the corresponding microwave billiard. Thus first intensity distributions of the electric field strength were measured with the so called perturbation body method \cite{Sridhar1992}. A small metallic body alters the resonance frequency $f_0$ of the cavity, where according to the Maier-Slater theorem \cite{MaierSlater1952} the frequency shift depends on the difference of the 
squares of the electric field strength $E$ and the magnetic field $B$ at its location inside the cavity.
\begin{equation}
\Delta f=f_0(c_1 E^2 -c_2 B^2).
\label{eqMaier}
\end{equation}
Here, $c_1$ and $c_2$ are constants determined by the geometry and material of the perturbation body. In order to obtain the electric field strength distribution, the 
perturber needs to be moved inside the closed resonator. This is best done 
with an external guiding magnet.
However, since we are interested in the electric field strength only, the magnetic field contribution in Eq.~(\ref{eqMaier}) has to be removed.  In earlier experiments 
\cite{Sridhar1992,Gokirmak1998,Hofferbert2000,sirko2004} with microwave billiards  the magnetic field contribution to the frequency shift was minimized by using a needle like metallic bead. We, for the first time, used a \emph{nonmetallic} perturber (of so called magnetic rubber) which is a rubber-like plastic combined with barium ferrite powder, whose grain size is of the same order of magnitude (1--10 $\mu \rm{m}$) as the skin depth of radio frequency waves. Therefore eddy currents cannot propagate, such that the magnetic rubber does not interact with the microwave magnetic field $B$. Accordingly there is no contribution to $\Delta f$ in Eq.~(\ref{eqMaier}).

We first performed test measurements with the rectangular cavity and a cylindrical perturbation body with a height of 5~mm and a diameter of 2~mm. Typically the positive frequency shift induced by the electric field strength is of the order of 100 kHz. As expected, we did not detect any negative frequency shifts, 
which would result from the magnetic field term in Eq.~(\ref{eqMaier}) within the experimental uncertainty of the frequency shift measurement of about 1 kHz. Moving the perturbation body  with an external guiding magnet across the billiard surface with a spatial resolution of about one-tenth of a
wavelength and measuring the frequency shifts $\Delta f$ at each of its positions yields the intensity distribution of the 
electric field strength in the whole microwave billiard. The shift $\Delta f$ is
determined as in \cite{Dembowski1999} via phase shift measurements. In the frequency list mode the vectorial network analyzer HP8510C allowed the simultaneous measurement of the phases at up to 30 different
frequencies. This enabled us to crucially reduce the measurement time, we measure 30 intensity distributions in one raster scan for a grid with 200$\times$100 points in 12 hours. To minimize the influence of a temperature drift, which causes an additional frequency shift of about 40 kHz/K, the temperature of the microwave billiard was stabilized to within 0.1 K.
\par Due to the symmetry of the barrier billiard there are two types of modes, anti-symmetric and symmetric ones with respect to reflections at the barrier, respectively. The former are trivial as they coincide with those of the rectangular billiard, the latter are exactly those we are interested in. 
We were able to measure, with the advanced experimental method described above, the intensity distributions of altogether 590 modes for level numbers $N=$90--680. In contrast to the technique described in \cite{Stoeck1992} with the perturbation body method we detect the $E$ field intensity only. For a detailed investigation of the symmetric mode properties we developed a special procedure to reconstruct the $E$ field, i.e.\ the quantum mechanical wave function $\psi$, itself and its sign from the measured intensity distribution.
\par
\begin{figure}[!ht]
{\includegraphics[width=\linewidth]{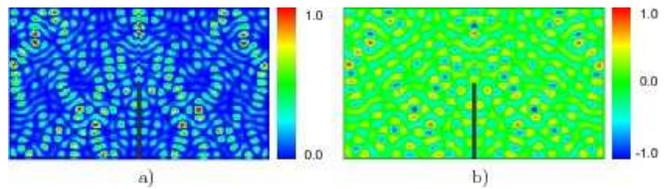}}
\caption{(Color online) (a) The measured intensity distribution for the mode $N=613$. (b) The reconstructed $E$ field, i.e.\ the wave function. The corresponding color scales are at r.h.s.\ of the graphs.} 
\label{figMeasReconstr}
\end{figure}

For this we fit the square of a sine series to the measured field intensity.
Along a line of constant $x$, $x=x_0$, in the billiard plane the latter can be represented as
\begin{equation}
\psi_k^2 (x=x_0,y)=\left[ \sum_{i=1}^n A_i(x_0) \sin k_{i}y\right]^{2} ,
\label{eqSinSer}
\end{equation}where $k_i=\pi i / l_y$, $l_y$ is the length of the billiard along the line $x=x_0$, and
$A_i(x_0)$ are the expansion coefficients. With this choice of $k_i$ the Dirichlet boundary
condition is automatically fulfilled. The number of terms giving a significant contribution to the sum can be estimated semi-classically, i.e.\ $n \sim 2 l_y/\lambda$ and $\lambda=2\pi/k$.
For the determination of the expansion coefficients $A_i(x_0)$ in Eq.~(\ref{eqSinSer}) a nonlinear fit routine which is based
on the Levenberg-Marquardt algorithm \cite{NumRecip} was used. The nonlinear fit requires a trial vector of coefficients as initial values. We used random numbers which are uniformly distributed in the interval from -$\mathrm{max}(\left|\psi(x=x_0,y)\right|)$ to $\mathrm{max}(\left|\psi(x=x_0,y)\right|)$ as entries of the trial vector, whose
typical dimension is about $n=20$. The nonlinear fit has no unique solution as the Levenberg-Marquardt algorithm
only provides a set of coefficients $A_i(x_0)$ which gives a local minimum in $\chi^2$. To find the global minimum we
generated about 1000 trial vectors and carried out a fit for each of these. Then that set of 
coefficients $A_i(x_0)$, which gives the smallest $\chi^2$, provides the wave function values along the line with constant $x=x_0$. 
Due to the continuity of the wave function for the neighboring lines 
we can use the so determined coefficients as entries of their trial vectors.  
An example for the measured intensity and the reconstructed field of the eigenmode with $N=613$ and frequency $f=5.48004~\rm{GHz}$ is shown, respectively, in Fig.~\ref{figMeasReconstr}.
 From 290 measured intensity distributions corresponding to the symmetric wave functions we  reconstructed 230. For the remaining 60 distributions, due to large noise or nearly overlapping resonances, reconstruction was not possible.
\par
\begin{figure}[!ht]
{\includegraphics[width=\linewidth]{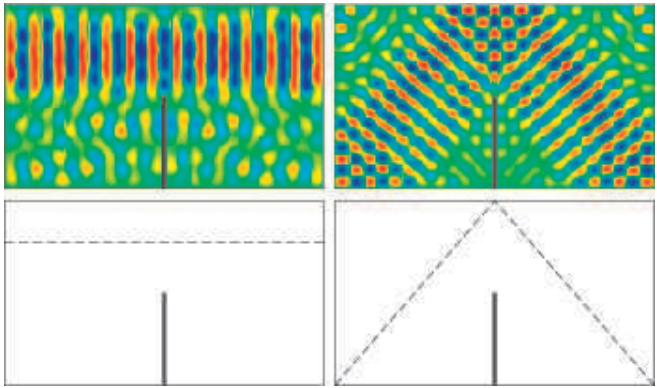}}
\caption{(Color online) The top row shows examples of experimentally obtained superscars in the barrier billiard. The color scale is the same as in Fig.~\ref{figMeasReconstr}(b). In the bottom row the corresponding classical orbits are indicated (dashed lines).} 
\label{figScarExampl0}
\end{figure}
\begin{figure}[!ht]
{\includegraphics[width=\linewidth]{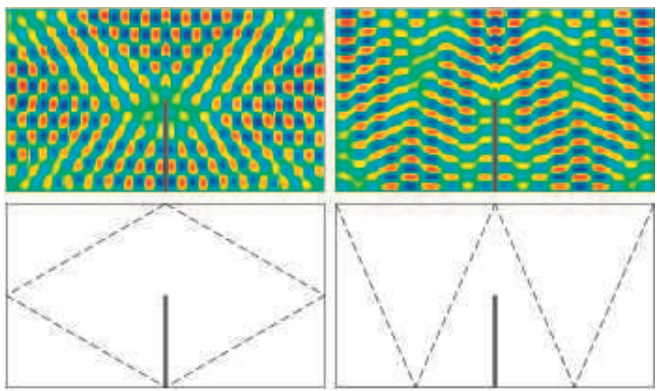}}

\caption{(Color online) The same as in Fig.~\ref{figScarExampl0} but for two other types of superscars.} 
\label{figScarExampl2}
\end{figure}

Among the large number of wave functions several superscarring ones were observed. Four examples of detected superscars are shown in Figs.~\ref{figScarExampl0}~and~\ref{figScarExampl2}. 
 Each wave function shows a clear structure which is obviously related to the corresponding periodic orbit (PO).
\par
\begin{figure}[!ht]
{\includegraphics[width=\linewidth]{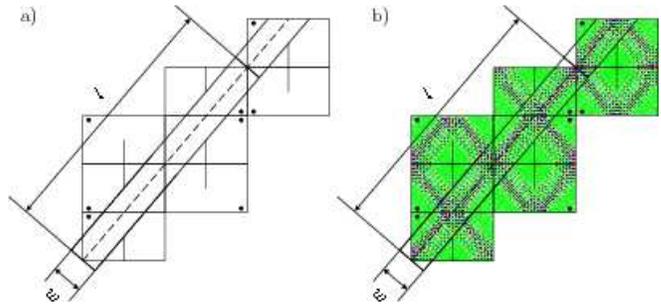}}
\caption{(Color online) (a) Illustration of the construction of the PO channel. The black point in the upper left corner of the original billiard fixes its initial orientation. (b) The unfolding of the experimentally obtained superscarring wave function with number $N=587$. It has 110 wave maxima along the POC and one perpendicular to it, corresponding to the quantum numbers $m=55$ and $n=1$, respectively.
} 
\label{figBilliard}
\end{figure}

To understand the quantum nature of the superscars we briefly outline how the PO families of a barrier billiard can be constructed. For this purpose we mark one corner of the barrier billiard with a black point
as shown in Fig.~\ref{figBilliard}(a), thus defining the initial orientation of the billiard. We follow the orbit of the propagating particle (dashed line in Fig.~\ref{figBilliard}(a)) and mirror the billiard each time the particle hits a side of the billiard. This procedure is repeated until the mirrored copy of the billiard resumes its initial orientation. All parallel trajectories with the same length form a family of POs. Each family corresponds to an infinitely long periodic orbit channel (POC). This POC is confined to the region between the two straight lines which connect the singularities, i.e.\ the barrier tip and its images. These lines are called
the singular diagonals (SD). In the semi-classical limit we can associate with each classical trajectory a wave.
 Waves travelling in a POC are scattered at an infinite array of images of the barrier tip.
At the SD the intensity of the scattered wave is proportional to $1/\sqrt{k}$, where $k$ is the wave momentum \cite{Bogomol2003}. Therefore, it tends to zero and fulfills the Dirichlet boundary condition there with increasing $k$. Accordingly, in the semi-classical limit the SD act 
as perfect mirrors such that in each POC quasi states can be constructed. These are called unfolded scar
states in \cite{Bogomol2004} and are given as 
\begin{equation}
\Psi_{m,n}^{scar}(\xi,\eta)=e^{-i \,k_m\, \xi}\cdot \sin\left( \frac{\pi n}{ w}\eta\right) \cdot \chi(\eta).
\label{eqScar}
\end{equation}Here $\xi$ is the coordinate along the POC and $\eta$ ($0<\eta< w $) the one perpendicular to
it, and $w$ is the channel width. The quantity $\chi(\eta)$ is the characteristic function of the POC 
($\chi(\eta)=1$ if $0<\eta<w$, and $\chi(\eta)=0$ otherwise). The frequency of the eigenfunction (Eq.~(\ref{eqScar})) of the POC is 
\begin{equation}
f_{m,n}=\frac{c}{2\pi}\sqrt{k_m^2+\left( \frac{\pi n}{w}\right)^2 }.
\label{eqScarEnergie}
\end{equation} Due to the Dirichlet boundary condition the wave gains a phase of $\pi$ each time the classical trajectory crosses the billiard boundary. If the PO experiences an even number of boundary crossings while travelling through the POC, then $k_m=2\pi\,m/l$, otherwise $k_m=\pi\,(2m+1)/l$, were $l$ is
the length of the PO. When folding back the unfolded superscar states of Eq.~(\ref{eqScar}), complicated expressions for folded scar states 
$\Psi_{m,n}(x,y)$ are obtained which, except on the SD, obey the Schr{\"o}dinger equation for the barrier billiard
$\left(\Delta +\left(\frac{2\pi}{c} f_{m,n}\right)^2\right)\Psi_{m,n}(x,y)=0$ with Dirichlet boundary conditions on the billiard boundary. As can be seen in Fig.~\ref{figBilliard}(b) the experimentally obtained superscarring wave functions are mainly concentrated inside the corresponding POC. This confirms the approximate validity of the ansatz for the superscar wave functions given in Eq.~(\ref{eqScar}).
\par
\begin{figure}[!ht]
{\includegraphics[width=\linewidth]{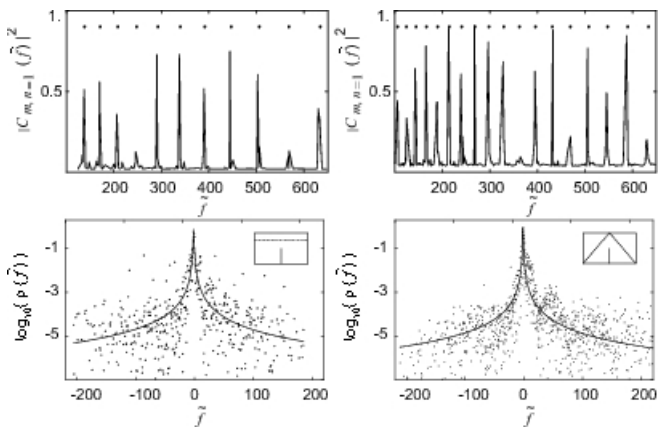}}

\caption{ The top row shows the overlap of measured wave functions of the barrier billiard with
analytically constructed scar states for different POs. The points indicate the predicted
positions of the superscar at the rescaled frequency $\tilde{f}$. In the bottom row the corresponding local density is plotted. The solid lines indicate the best fit of a Breit-Wigner shape to the data. The insets show the
corresponding PO families.}
\label{figUberlap}
\end{figure}

 A typical eigenfunction of the barrier billiard may have contributions from many scar states. In order to obtain a quantitative measure for this we followed Ref.~\cite{Bogomol2004} and computed 
the overlap integral of folded scar states $\Psi_{m,n}^{scar}(x,y)$ given in Eq.~(\ref{eqScar}) with the measured wave function
$\Psi_{\tilde f_\lambda}(x,y)$ at the unfolded resonance frequencies $\tilde f_\lambda$, where the unfolding was performed as usual \cite{Stoeckmann:buch00,richter97} using the Weyl formula \cite{Weyl1912},
\begin{equation}
C_{m,n}(\tilde{f}_{\lambda})=\int\Psi_{m,n}^{scar}(x,y)\Psi_{\tilde{f}_{\lambda}}(x,y)\,{\rm d}x\,{\rm d}y\,.
\label{eqOvelap}
\end{equation}
In the computations we fixed the number of excitations perpendicular to the POC to $n=1$; $m$ is obtained 
from the condition that $|\tilde{f}_{\lambda}-\tilde{f}_{m,n}|$ should be minimal. The resulting squares of
overlap coefficients $|C_{m,n}(\tilde{f})|^2$, versus the unfolded resonance frequency $\tilde{f}$ for two different families of superscars, are shown in Fig.~\ref{figUberlap}. The frequencies of the superscars are given by Eq.~(\ref{eqScarEnergie}). Near almost all predicted superscar frequencies, marked by points in Fig.~\ref{figUberlap}, we found a wave function with a strong contribution to the integral of overlap with a constructed superscar state.
\par To analyze the overlap of the constructed superscars with the neighboring non-scarred states, we calculated the average over all the spikes showed in Fig.~\ref{figUberlap}, and obtained the so-called local density of states averaged over different $m$,
\begin{equation}
\rho_{n}(\tilde{f})=\bigg\langle\sum_{\lambda}|C_{m,n}(\tilde{f}_{\lambda})|^2\delta(\tilde{f}-\tilde{f}_{\lambda}+\tilde{f}_{m,n})\bigg\rangle_{m}.
\label{eqRho}
\end{equation}
In the bottom row of Fig.~\ref{figUberlap} this quantity is shown for the two scar families in the top row of the figure. It is evident that the constructed superscars Eq.~(\ref{eqScar}) have a certain overlap with the neighboring states. 
The finite width of the local density curves indicates that the Dirichlet boundary condition along the SD is an approximation, that is, the field can leak out of the POC. 
As can be seen in Fig.~\ref{figUberlap}, it seems to be reasonable to approximate the averaged local density of the superscar states by a Breit-Wigner shape, but further work on this is in progress. 

\par In summary, we have measured the electric field strength distributions at 590 resonance frequencies of the barrier billiard and reconstructed almost 
all corresponding symmetric wave functions. Many of the wave functions show a surprisingly pronounced structure of so called superscars which is connected with classical 
PO families. We plan to extend our measurements of wave functions in barrier billiards with one or more non-symmetrically placed barriers. The statistical properties of the nodal domains of polygonal billiards are another very interesting problem, which will be investigated.
\par
This work has been supported within the DFG grant SFB634.

\end{document}